# Comment on arXiv:2210.01114
# Optical Saturation Produces Spurious Evidence for Photoinduced Superconductivity in $K_3C_{60}$

M. Buzzi[1,*], D. Nicoletti[1], E. Rowe[1], E. Wang[1], and A. Cavalleri[1,†]

[1]*Max Planck Institute for the Structure and Dynamics of Matter, Hamburg, Germany*
*\*michele.buzzi@mpsd.mpg.de*
*†andrea.cavalleri@mpsd.mpg.de*

In the manuscript arXiv:2210.01114, Dodge and co-authors discuss the influence of pump-probe profile deformations on the reconstructed non-equilibrium optical conductivity of $K_3C_{60}$. They state that when pump-induced saturation of the probe response is taken into account, the reconstructed optical properties are not superconducting-like, as was claimed in a number of experimental reports by our group. We show here that the conclusion reached by Dodge *et al.* is unjustified. In fact, independent of the specific model, including the problematic saturation profile proposed by the authors, the reconstructed optical properties are those of a finite temperature superconductor. The true fingerprint of superconductivity, which is the $1/\omega$ divergence of the imaginary conductivity, $\sigma_2(\omega)$, is retained and is virtually independent of the chosen model. The only model-dependent feature is the degree of "gapping" in $\sigma_1(\omega)$. In all cases the extracted optical properties reflect the presence of residual quasiparticles, which at finite temperatures are inevitably present alongside the superfluid.

In reference (*1*), Dodge and co-authors critique the determination of the transient optical properties of photo-excited $K_3C_{60}$, which in a series of recent papers (*2-4*) were reported to be strongly reminiscent of superconductivity.

The issue being debated here is the following. In mid-infrared (MIR) pump - terahertz (THz) probe measurements on $K_3C_{60}$, the probe field penetrates deeper into the material than the pump. Hence, the raw experimental data contains a signal averaged over excited and unexcited regions of the sample, resulting in an underestimated response. To extract the transient optical conductivity of the photo-excited material, one must account for this effect.

The pump and probe are attenuated inside the material according to their respective Beer-law extinction depths. In early papers (*2-4*), a linear dependence of the light induced changes in the THz response on MIR pump fluence was assumed. The uncertainty introduced by this approximation was already addressed in Ref. (*2*), and it was concluded that for changes in penetration depth mismatch up to 25%, the reconstructed response would not be appreciably affected, at least in terms of the qualitative nature of the observed non-equilibrium state. In more recent work (*5*) data reconstructed assuming a linear dependence of the pump induced changes on the pump electric field (i.e. square-root with fluence) were reported, and shown to be superconducting-like.

Dodge *et al.* choose a model in which they assume that the changes in the THz optical properties saturate above a certain fluence. Implicit in this assumption is that the material must be undergoing a photo-induced transition into a different phase. This phenomenon is later questioned by the authors, making this argument in itself not entirely consistent and hence problematic. Using this model, they re-evaluate the photo-induced changes in the optical conductivity, concluding that the response is not superconducting-like. Here, we show that these assertions are unjustified.

This is best appreciated by inspecting Fig 1 in this comment. Therein, the equilibrium optical spectra measured by Degiorgi et al. (6) in $K_3C_{60}$ single crystals below and above $T_C$ = 20 K are compared to those obtained for the transient state after reconstruction. The normal state spectra measured in the crystals (panel a) and in the pellets (red curves in panels b, c, d) differ in absolute value and shape of the conductivity. This is well understood as a consequence of a different carrier density (see discussion in Ref. (2)). However, this difference is not important for the following discussion.

Figure 1 demonstrates that regardless of the model used for reconstruction, and even in the sub-optimal conditions (3) reported in panel (b), the response is superconducting-like and bears a strong similarity to the equilibrium optical properties below $T_C$. More specifically, *the* characteristic feature of superconductivity, that is the $1/\omega$ divergence of the imaginary part of the conductivity, $\sigma_2(\omega)$, is always retained. The degree of "gapping" observed in $\sigma_1(\omega)$ changes depending on the model, and even in the most unflattering assumption it is reminiscent of the "gapping" observed in equilibrium (panel a) for T = 15 K (~0.7 $T_C$).

In panels (c) and (d), we show more recent results acquired with optimized fluence and excitation wavelength (4, 5). There, the "gapping" in $\sigma_1(\omega)$ is even more pronounced. Note that the regimes where we observe negative $\sigma_1(\omega)$, indicative of amplification of the probe light, is topic for a separate discussion (5).

The experiments reported here are also flanked by a number of other studies of the same physical phenomenon using other techniques, which show vanishing two-probe electrical resistance (3), superconducting-like pressure scaling of the transient photo-conductivity (7) and nonlinear electrical properties (8).

Finally, a series of experiments performed in charge transfer salts (9, 10), in which similar MIR excitation was combined with THz probing, yielded clear signatures of a fully gapped superconductor in a situation without pump-probe penetration depth mismatch.

In summary, we have shown how the response of photoexcited $K_3C_{60}$ reported in Refs. (2-5, 7) is at the very least "superconducting-like", regardless of the assumption used to extract the transient optical spectra. We believe that the criticism raised by Dodge et al. (1) has little merit.

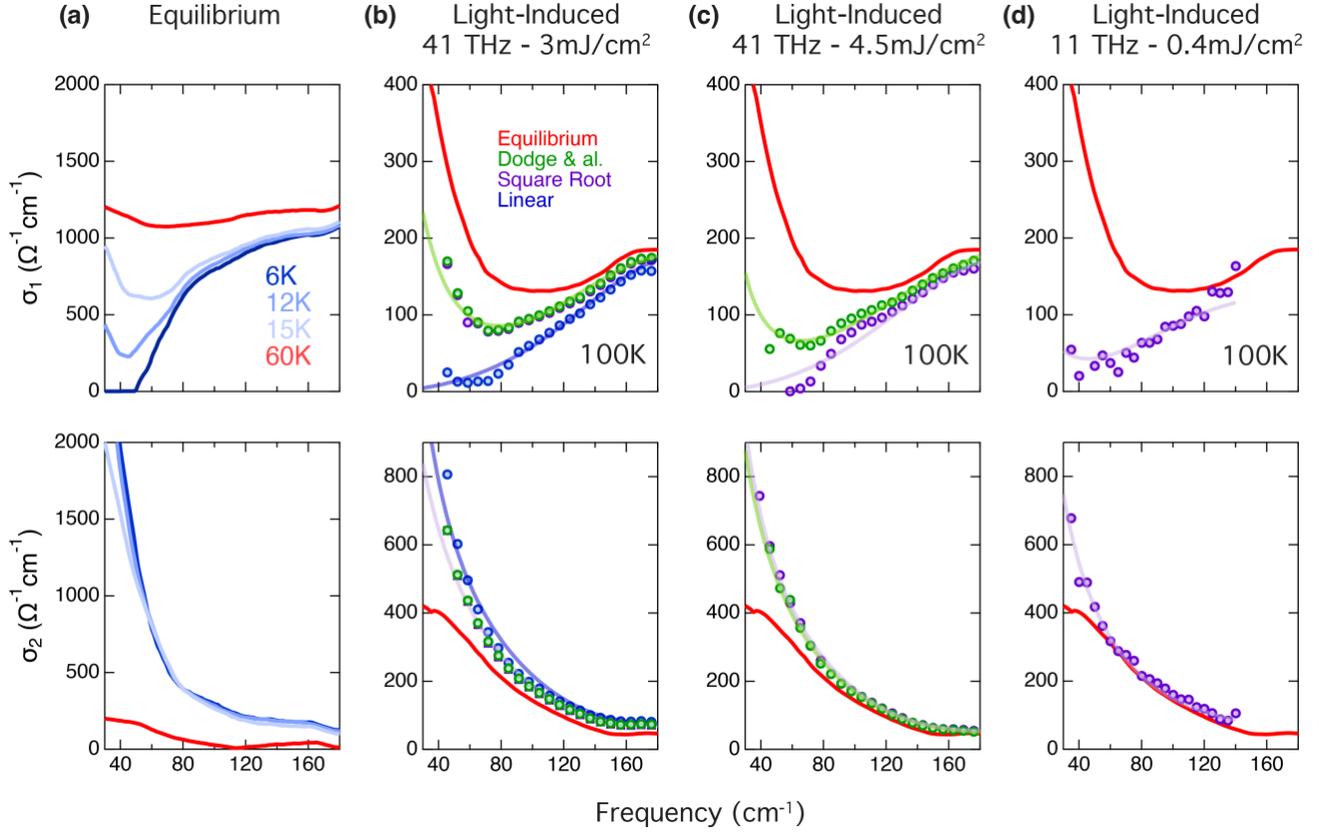

**Figure 1:** **(a)** Complex optical conductivity, $\sigma_1(\omega) + i\sigma_2(\omega)$, of $K_3C_{60}$ single crystals measured at different temperatures across the equilibrium superconducting transition. These data have been obtained via Kramers-Kronig transformations from the reflectivity spectra in Ref. (*6*). **(b)** Same quantities as in (a), measured in pressed $K_3C_{60}$ powders at equilibrium (red solid curves) and at 1 ps time delay after photo-excitation with 3 mJ/cm² fluence (full circles). Different colours refer to different models used to describe the pump-probe penetration depth mismatch: a model assuming linear fluence dependence (blue) as that used in Refs. (*2-5, 7*), a square-root model (purple, as described in Ref. (*5*)), and the saturation model proposed by Dodge *et al.* (green, see Ref. (*1*)). Data reproduced from Refs. (*1, 3*). **(c)** Same quantities as in (b) but measured at a higher excitation fluence of 4.5 mJ/cm². Here, nearly full $\sigma_1(\omega)$ gapping is achieved even with a square-root (purple) or saturation (green) model. Data reproduced from Refs. (*1, 4*). **(d)** Same quantities as in (b,c) but measured at a different excitation frequency of 11 THz, fluence of 0.4 mJ/cm², and 11.5 ps pump probe delay. Data reproduced from Ref. (*5*) and not analysed in Ref. (*1*). Solid lines are fits to the transient optical spectra with a two-fluid model. All transient spectra in (b-d) were measured at T = 100 K.